\documentclass[aps,prd,groupedaddress,showpacs,showkeys]{revtex4}
\usepackage{epsfig}

\newcommand{\eq}{\begin{eqnarray}}
\newcommand{\en}{\end{eqnarray}}

\begin{document}

\title{On the two-photon decay width of the sigma meson}
\author{ Francesco Giacosa$^1$, Thomas Gutsche$^2$, Valery E. Lyubovitskij$%
^2 $\thanks{
On leave of absence from the Department of Physics, Tomsk State University,
634050 Tomsk, Russia} \vspace*{1.2\baselineskip}}
\affiliation{$^1$ Institut f\"ur Theoretische Physik, Johann Wolfgang
Goethe-Universit\"at, Max von Laue-Strasse 1, 60438 Frankfurt, Germany 
\vspace*{0.3\baselineskip}\\
$^2$ Institut f\"ur Theoretische Physik, Universit\"at T\"ubingen, \\
Auf der Morgenstelle 14, D-72076 T\"ubingen, Germany \vspace*{%
0.3\baselineskip}\\
}

\begin{abstract}
We shortly report on the two-photon decay width of the light $\sigma$-meson
interpreted as a quarkonium state. Results are given in dependence on the $%
\sigma$-mass and the constituent mass of the light quark. The triangle
quark-loop diagram, responsible for the two-photon transition, is carefully
evaluated: a term in the transition amplitude, often omitted in literature,
results in destructive interference with the leading term. As a result we
show that the two-photon decay width of the $\sigma $ in the quarkonium
picture is less than 1 keV for the physical range of parameters.
\end{abstract}

\maketitle

\section{Introduction}

The two-photon decay of scalar mesons represents a valuable mechanism to
possibly pin down their internal structure (see~\cite%
{amslerrev,achasov,li,efimov,pennington,closekirk,volkov,faessler,giacosa1,dewitt,branz,scadron1,scadron2,schumacher,narison}
and references therein). In particular, the transition of the
scalar-isoscalar resonance $\sigma \equiv f_{0}(600)$ into $\gamma \gamma $
has received much attention in the literature. It is commonly believed that
a decay width of about 3 - 5 keV would favor a quarkonium interpretation of
the $f_{0}(600).$ In this short work we aim to show that this is not the
case: the decay width of a scalar-isoscalar quark-antiquark state, with
flavor wave-function $\overline{n}n=\sqrt{\frac{1}{2}}(\overline{u}u+%
\overline{d}d)$ and a mass between 0.4 and 0.8 GeV as favored by recent
studies (a mass of about 0.44 GeV is the outcome of \cite{caprini}), turns
out to be smaller than 1 keV for the physical range of parameters. When
evaluating the related quark triangle-loop diagram of Fig. 1 care has to be
taken concerning gauge invariance, for a comprehensive and detailed analysis
we refer to \cite{faessler}: a (often neglected) term generating a
consistent suppression of the decay amplitude is present, as will be
discussed in Sections II and III . The omission of this term generates an
overestimate of the two-photon decay rate by at least a factor of $2.25.$
Considering the relevance of this process related to the nature of the $%
\sigma $ meson, and in more general of scalar mesons (see for instance \cite%
{scalars} and Refs. therein), we consider it as important to stress this
point for future considerations about the interpretation of the enigmatic $%
\sigma $-resonance.

\begin{figure}[btbp]
\centering{\ }
\par
\vspace*{-.5cm} \epsfig{figure=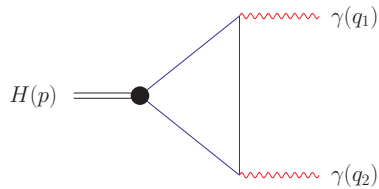,scale=.5} \vspace*{-.35cm}
\caption{Quark-loop diagram contributing to $H \protect\gamma \protect\gamma 
$ decay, where $H = \protect\pi$ or $\protect\sigma$.}
\end{figure}

The triangle quark loop diagram of Fig. 1 is typical for theories with
quarks as effective degrees of freedom \cite%
{efimov,volkov,faessler,giacosa1,scadron1,scadron2,hatsuda,klevanski}. It is
evaluated both in the framework of local and nonlocal $\sigma $-$\overline{n}%
n $ vertices. In the local case the Goldberger-Treiman relation on the quark
level and the linear realization of chiral symmetry allow to fix the
corresponding $\sigma $-$\overline{n}n$ coupling constant. In the nonlocal
case the finite size of the $\sigma $-meson interpreted as a quarkonium
state is described by means of a covariant vertex function. The results of
local and nonlocal approaches are similar when $M_{\sigma }$ is sufficiently
below threshold set by the sum of constituent quark masses. Close to
threshold care has to be taken in the local case since a second possible
problem can arise: the explicit momentum dependence of the $\sigma $-$%
\overline{n}n$ coupling constant cannot be neglected further. This is also a
delicate point to be treated with attention.

The present article is organized as follows: in the next two sections we
elaborate on the formalism and the results for $\sigma \rightarrow \gamma
\gamma $ as based on a local and nonlocal interaction Lagrangian,
respectively. In section IV we summarize and give our conclusions.

\section{Local case}

We consider the following local (L) interaction Lagrangian 
\begin{equation}
\mathcal{L}_{\mathrm{int}}^{\text{L}}(x) \,=\, \frac{g_{\sigma }}{\sqrt{2}}
\sigma (x)\,\bar{q}(x)q(x) + \frac{g_{\pi }}{\sqrt{2}}\bar{q}(x) i\gamma_{5} 
\vec{\pi}(x) \vec{\tau} q(x)  \label{intloc}
\end{equation}%
where $q^{T} = (u,d)$ is the quark doublet of $u$ and $d$ quarks with the
constituent mass $m_{q}=m_{u}=m_{d}$ (we restrict to the isospin limit) to
be varied between 0.25 and 0.45 GeV, $\sigma (x)$ and $\vec{\pi}(x)$
represent the scalar-isoscalar quarkonium and the isotriplet pion field,
respectively, $g_{\sigma }$ and $g_{\pi }$ are the corresponding coupling
constants (which are later related via symmetry and low-energy
considerations). We will denote the meson masses by $M_{\pi} = M_{\pi^0} =
134.9766$ MeV and $M_{\sigma }$, respectively. The latter will be varied
between 0.4 and 0.8 GeV.

The decay of $H=\pi ^{0},\sigma $ into $\gamma \gamma $ is obtained by
evaluating the diagram of Fig. 1. The decay width is explicitly given by: 
\begin{equation}
\Gamma _{H\rightarrow \gamma \gamma }=\frac{\pi }{4}\,\alpha
^{2}\,g_{H\gamma \gamma }^{2}\,M_{H}^{3}\,,\ \ \ \ \text{{}}H=\pi
^{0},\sigma \,,  \label{hgg}
\end{equation}%
where $g_{H\gamma \gamma }=g_{H}N_{c}Q_{H}I_{H}/(2\pi ^{2})$ is the
effective $H\gamma \gamma $ coupling constant, $\alpha $ is the fine
structure constant, $N_{c}=3$ the number of colors and $g_{H}$ refers to the
coupling constant $g_{\pi }$ or $g_{\sigma }$ entering in the interaction
Lagrangian of Eq. (\ref{intloc}). The charge factors $Q_{\pi ^{0}}=\frac{1}{%
\sqrt{2}}(\frac{4}{9}-\frac{1}{9})=\frac{3}{9\sqrt{2}}$ and $Q_{\sigma }=%
\frac{1}{\sqrt{2}}(\frac{4}{9}+\frac{1}{9})=\frac{5}{9\sqrt{2}}$ correspond
to the flavor wave functions $\pi ^{0}\equiv \sqrt{\frac{1}{2}}(\overline{u}%
u-\overline{d}d)$ and $\sigma \equiv \sqrt{\frac{1}{2}}(\overline{u}u+%
\overline{d}d)$. Finally, the loop integrals $I_{H}$ \cite{faessler}
corresponding to Fig. 1 are functions of $m_{q}$ and $M_{H}$, which are
explicitly given by 
\begin{eqnarray}
I_{\pi ^{0}} &=&I_{\pi ^{0}}(m_{q},M_{\pi })=m_{q}\int_{0}^{1}\mathrm{d}^{3}%
\mathrm{\alpha }\,\delta \left( 1-\sum_{i=1}^{3}\alpha _{i}\right) \frac{1}{%
m_{q}^{2}-M_{\pi }^{2}\alpha _{1}\alpha _{2}}=\frac{2m_{q}}{M_{\pi }^{2}}%
\arcsin ^{2}\left( \frac{M_{\pi }}{2m_{q}}\right) ,  \label{ipion} \\
I_{\sigma } &=&I_{\sigma }(m_{q},M_{\sigma })=m_{q}\int_{0}^{1}\mathrm{d}^{3}%
\mathrm{\alpha }\,\delta \left( 1-\sum_{i=1}^{3}\alpha _{i}\right) \frac{%
1-4\alpha _{1}\alpha _{2}}{m_{q}^{2}-M_{\sigma }^{2}\alpha _{1}\alpha _{2}}=%
\frac{2m_{q}}{M_{\sigma }^{2}}\left[ 1+\left( 1-\frac{4m_{q}^{2}}{M_{\sigma
}^{2}}\right) \arcsin ^{2}\left( \frac{M_{\sigma }}{2m_{q}}\right) \right]
\,.  \label{isigma}
\end{eqnarray}%
Note that the only difference between $I_{\pi ^{0}}$ and $I_{\sigma }$ is
the term proportional to $-4\alpha _{1}\alpha _{2}$ present in the integral
expression of $I_{\sigma }$, which is generally neglected in the literature
(that is, the amplitudes of $\pi ^{0}\rightarrow \gamma \gamma $ and $\sigma
\rightarrow \gamma \gamma $ cannot be set equal to each other as done, for
instance, in~\cite{scadron1,scadron2}). The presence of the term $-4\alpha
_{1}\alpha _{2}$ generates a destructive interference with the first term,
which leads to a sizable reduction of the full amplitude $I_{\sigma }.$
Quantitatively the ratio of amplitudes is limited by $I_{\sigma
}(m_{q},x)/I_{\pi ^{0}}(m_{q},x)<0.667$ for values of $0<x<2m_{q}$ in the
region of applicability. Thus, neglecting the additional term in $I_{\sigma
} $ implies an overestimate of the decay rate $\Gamma _{\sigma \rightarrow
\gamma \gamma }$ by at least a factor of $0.667^{-2}=2.25$, as already
indicated in the Introduction. Notice that we compare the decay amplitudes $%
I_{\sigma }(m_{q},x)$ and $I_{\pi ^{0}}(m_{q},x)$ but not the corresponding
decay widths: as shown below these will differ consistently because of the
dependence on the third power of the meson mass in eq. (\ref{hgg}).

Let us now turn to the explicit calculation of decay rates. The
Goldberger-Treiman (GT) relation $g_{\pi } = m_{q} \sqrt{2}/F_{\pi }$ with $%
F_{\pi }=92.4$ MeV allows to determine $g_{\pi }.$ As an outcome we obtain
for the decay width $\Gamma_{\pi^{0}\rightarrow \gamma \gamma }=7.73 - 8.12$
eV for constituent quark masses in the range $m_{q}=0.45 - 0.25$ GeV, in
good agreement with the experimental result $\Gamma
_{\pi^{0}\rightarrow\gamma \gamma }^{\text{exp}} = 7.7\pm 0.5\pm 0.5$ eV 
\cite{pdg}. Only a very weak dependence on $m_{q}$ is observed.

The linear realization of chiral symmetry implies $g_{\sigma }=g_{\pi }$ 
\cite{faessler,scadron1,klevanski}. Predictions for $\Gamma _{\sigma
\rightarrow \gamma \gamma }$ can then be obtained in dependence on the
effective quark mass $m_{q}$ and on $M_{\sigma }.$ Note that we limit the
parameter space by the relation $M_{\sigma }<2m_{q}$: in fact, only when
this condition is met the amplitude $I_{\sigma }$ remains real and no
unphysical decay of the sigma meson into a quark-antiquark pair is included.
Furthermore, the condition $g_{\sigma }=g_{\pi }$ can only be employed, if $%
M_{\sigma }$ is safely below the threshold $2m_{q}.$ For $M_{\sigma }\sim
2m_{q}$ the momentum dependence of $g_{\sigma }$ becomes non-negligible
leading to a value for $g_{\sigma }$ smaller than the one obtained in the GT
limit $m_{q}\sqrt{2}/F_{\pi }$; in the next section we illustrate this point
in the context of the nonlocal approach.

\begin{center}
\bigskip

\textbf{Table 1}: $\Gamma _{\sigma \rightarrow \gamma \gamma }$ in the local
case for

$m_{q}=0.25 - 0.45$ GeV at $M_{\sigma }=0.440$ GeV

\vspace*{.1cm}

\begin{tabular}{|l|l|l|l|l|l|}
\hline
$m_{q}$ (GeV) & 0.25 & 0.3 & 0.35 & 0.4 & 0.45 \\ \hline
$\Gamma _{\sigma \rightarrow \gamma \gamma }$ (keV) & 0.54 & 0.45 & 0.41 & 
0.39 & 0.37 \\ \hline
\end{tabular}
\end{center}

In Table 1 we report the results for $\Gamma _{\sigma \rightarrow \gamma
\gamma }$ at a fixed pole mass of $M_{\sigma }=440$ MeV as favored by recent
theoretical and experimental works~\cite{caprini,pdg}. The results are
weakly dependent on $m_{q}$ and clearly point to a decay width smaller than
1 keV, when the sigma meson is interpreted as a quarkonium state. Note for
example that the omission of the previously discussed term in Eq. (\ref%
{isigma}) implies an overestimated decay width of $\Gamma _{\sigma
\rightarrow \gamma \gamma }=1.18$ keV for a value of $m_{q}=0.3$ GeV, to be
compared to the correct result of $0.49$ keV reported in Table~1.

In Fig. 2 we indicate the dependence of $\Gamma _{\sigma \rightarrow \gamma
\gamma }$ on $M_{\sigma }$ for values of the constituent quark mass, $%
m_{q}=0.35$ and $0.4$ GeV, very often used in phenomenological studies. For
values of $M_{\sigma }$ safely below threshold (up to $0.5$ GeV) results for 
$\Gamma _{\sigma \rightarrow \gamma \gamma }$ lie below 1 keV and
essentially do not depend on the quark mass. For values of $M_{\sigma }$
approaching threshold the dependence on $m_q$ becomes more pronounced, where
the results for $\Gamma _{\sigma \rightarrow \gamma \gamma }$ eventually
grow beyond 1 keV. However, the local approach is no longer applicable for
values of $M_{\sigma}$ close to threshold, as will be evident from the
discussion of the next section.

\begin{figure}[tp]
\vspace*{-.5cm} \epsfig{figure=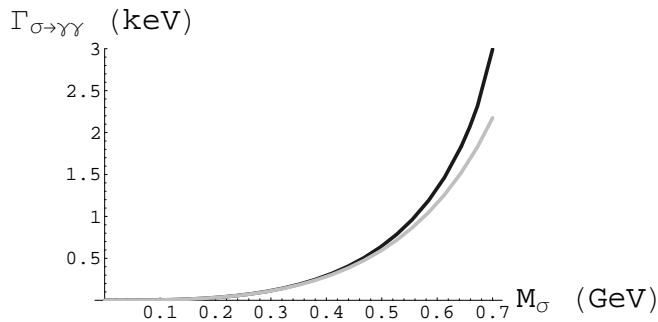,scale=.75} \vspace*{-.5cm} \centering%
{\ }
\caption{$\Gamma _{\protect\sigma \rightarrow \protect\gamma \protect\gamma}$
in the local case as function of $M_{\protect\sigma }$ for $m_{q}=0.35$
(dark) and $m_{q}=0.4 $ (gray).}
\end{figure}

\section{Nonlocal case}

Next we the study the sigma meson described by the nonlocal (NL) interaction
Lagrangian \cite{faessler} 
\begin{equation}
\mathcal{L}_{\mathrm{int}}^{\text{NL}}(x)\,=\,\frac{g_{\sigma }}{\sqrt{2}}
\sigma (x)\,\int d^{4}y\,\Phi (y^{2})\,\bar{q}(x+y/2)q(x-y/2)\,,
\label{lnonloc}
\end{equation}
where the delocalization takes account of the extended nature of the
quarkonium state by the covariant vertex function $\Phi(y^{2})$. The
(Euclidean) Fourier transform of this vertex function is taken as $%
\widetilde{\Phi }(k_{E}^{2})=\exp (-k_{E}^{2}/\Lambda^{2}),$ also assuring
UV-convergence of the model. The cutoff parameter $\Lambda $ will be varied
between $1$ and $2$ GeV, corresponding to an extension of the $\sigma$ of
about $l\sim 1/\Lambda \sim 0.5$ fm. Previous studies~\cite{anikin} have
shown that the precise choice of $\widetilde{\Phi }(k_{E}^{2})$ affects only
slightly the result, as long as the function falls of sufficiently fast at
the energy scale set by $\Lambda $. The coupling $g_{\sigma }$ is determined
by the so-called compositeness condition $Z_{\sigma }=1-\Sigma _{\sigma
}^{\prime }(M_{\sigma }^{2})=0$ \cite{efimov,faessler,weinberg}, where $%
\Sigma _{\sigma }^{\prime }$ is the derivative of the $\sigma $-meson mass
operator given by 
\begin{equation}
\Sigma _{\sigma }(p^{2})=-g_{\sigma }^{2}N_{c}\int \frac{d^{4}k}{(2\pi )^{4}i%
} \,\widetilde{\Phi }^{2}(-k^{2})\,\mathrm{tr}\left[
S_{q}(k+p/2)S_{q}(k-p/2) \right] \, ,  \label{gsigma}
\end{equation}
where $S_{q}(k)=(m_{q}-\not\! k)^{-1}$ is the quark propagator. Note, the
compositeness condition is equivalent to the hadron wave function
normalization condition in quantum field approaches based on the solution of
the Bethe-Salpeter/Faddeev equation~\cite{roberts}. At this level it is
clear that $g_{\sigma }$ is a function of $M_{\sigma }$. In Fig. 3 we give
the dependence of $g_{\sigma }(M_{\sigma }^{2})$ at $m_{q}=0.35$ GeV for
cut-off values of $\Lambda =1$ and $2$ GeV, respectively, and indicate the
local limit with $g_{\sigma}=m_{q}\sqrt{2}/F_{\pi }.$ For low values of $%
M_{\sigma }$ the coupling $g_{\sigma }(M_{\sigma }^2)$ is a slowly varying
function, values of which for $\Lambda =1-2$ GeV also include the GT limit.
However, for values of $M_{\sigma}$ approaching threshold $g_{\sigma
}(M_{\sigma }^2)$ decreases below the local result (see details in Ref. \cite%
{pagliara}).

\begin{figure}[tp]
\centering{\ }
\par
\vspace*{-.5cm} \epsfig{figure=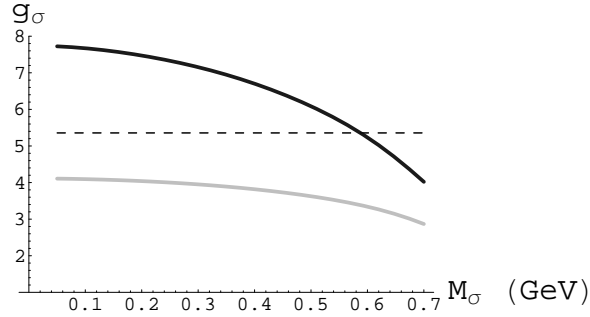,scale=.75} \vspace*{-.5cm}
\caption{$M_\protect\protect\sigma$-dependence of the coupling $g_{\protect%
\sigma }(M_{\protect\sigma }^{2})$ at $m_{q}=0.35$ for cut-off values of $%
\Lambda =1 $ (dark) and $2$ GeV (gray). The dashed line corresponds to the
GT limit.}
\end{figure}

We turn to the $\sigma \to \gamma \gamma $ decay amplitude, where a similar
suppression is found. Due to the presence of the vertex function $\Phi
(y^{2})$ inclusion of the electromagnetic interaction is achieved by gauging
the nonlocal interaction Lagrangian (\ref{lnonloc}): in addition to the
photon-quark coupling, already present in the local case, in leading order a
new vertex arises, where the photon couples directly to the $\sigma \gamma
\gamma $ interaction vertex, see \cite{faessler} for details. In particular,
in addition to the triangle diagram of Fig. 1 we have additional diagrams
(see Fig. 5 in Ref.~\cite{faessler}) to fully guarantee gauge invariance of
the transition amplitude. In practice it is convenient to split the
contribution of each diagram into a part which is gauge invariant and one
which is not. The remaining terms, which are not gauge invariant, cancel
each other in total and in the further calculation one should only proceed
with the gauge invariant terms of the separate diagrams. It was shown~\cite%
{faessler}, that the by far dominant contribution comes from the gauge
invariant part of the triangle diagram of Fig. 1. The gauge invariant parts
of the other diagrams are strongly suppressed (see discussion in Refs.~\cite%
{faessler,giacosa1}).

Following \cite{faessler,giacosa1} the contribution of the gauge-invariant
part of the triangle diagram to the two-photon decay width is given by: 
\begin{eqnarray}
\Gamma _{\sigma \rightarrow 2\gamma } &=&\frac{\pi }{4}\alpha ^{2}M_{\sigma
}^{3}\left[ \frac{g_{\sigma }}{2\pi ^{2}}Q_{\sigma }N_{c}I_{\sigma }\right]
^{2},\text{ }I_{\sigma }=I_{\sigma }^{(1)}+I_{\sigma }^{(2)}\,,
\label{GammaNL} \\
I_{\sigma }^{(1)} &=&m_{q}\int \frac{d^{4}k}{\pi ^{2}i}\,\widetilde{\Phi }
(-q^{2})\,\frac{1}{
(m_{q}^{2}-p_{1}^{2})(m_{q}^{2}-p_{2}^{2})(m_{q}^{2}-p_{3}^{2})}\,,
\label{isigma1} \\
I_{\sigma }^{(2)} &=& - m_{q}\int \frac{d^{4}k}{\pi ^{2}i}\,\widetilde{\Phi }
(-q^{2})\,\frac{\displaystyle{\frac{4}{M_{\sigma }^{2}}k^{2}-\frac{32}{
M_{\sigma }^{4}}}(kq_{1})(kq_{2})}{
(m_{q}^{2}-p_{1}^{2})(m_{q}^{2}-p_{2}^{2})(m_{q}^{2}-p_{3}^{2})}\,.
\label{isigma2}
\end{eqnarray}%
where $q_{1}$ and $q_{2}$ are the photon momenta and $p_{1}=k+q_{1},$ $%
p_{2}=k,$ $p_{3}=k-q_{2},$ $q=(p_{1}+p_{3})/2.$ The term $I_{\sigma }^{(2)}$
contributes with opposite sign relative to $I_{\sigma }^{(1)}$ leading to
destructive interference. In the local limit, i.e. $\Lambda \rightarrow
\infty $, $I_{\sigma }^{(2)}$ reduces to the term proportional to $-4\alpha
_{1}\alpha _{2}$ in (\ref{isigma}). Note that in the pion case only a term
analogous to $I_{\sigma }^{(1)}$ contributes.

\begin{figure}[bp]
\centering{\ }
\par
\epsfig{figure=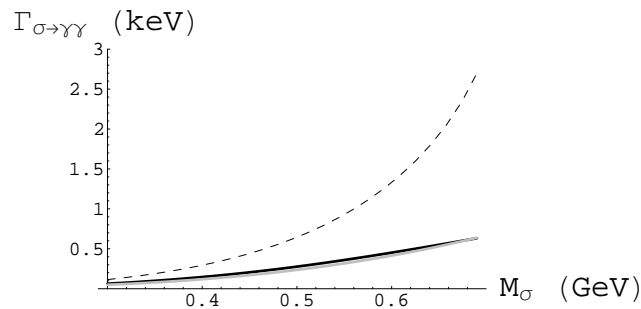,scale=.75} \vspace*{-.5cm}
\caption{$\Gamma _{\protect\sigma \rightarrow \protect\gamma \protect\gamma %
} $ in the nonlocal case as function of $M_{\protect\sigma }$ for $\Lambda
=1 $ GeV (dark) and $2$ GeV (gray). The quark mass is set to $m_{q}=0.35$.
The upper dashed line corresponds to the local limit evaluated in Section
II. }
\end{figure}

In Fig. 4 we report the results for $\Gamma _{\sigma \rightarrow
\gamma\gamma }$ in the nonlocal case as function of $M_{\sigma }$ for $%
m_{q}=0.35$ GeV, taking values of $\Lambda =1$ and $2$ GeV. We also indicate
the previous local result. While for small $M_{\sigma}$ both approaches,
local and nonlocal, agree, for increasing $M_\sigma$ the nonlocal approach
delivers smaller decay rates than the local counterpart, because of the
threshold effects described above. The nonlocal results depend very weakly
on the value $\Lambda$, implying that the numerical values for $%
\Gamma_{\sigma \rightarrow \gamma\gamma }$ are hardly model dependent. In
Table~2 we summarize our results for the two pole masses, $M_{\sigma }=0.44$
and $0.6$ GeV, choosing different values of $m_q$ both for $\Lambda =1$ GeV
and, in parenthesis, for $\Lambda =2$ GeV.

\begin{center}
\textbf{Table 2: } $\Gamma _{\sigma \rightarrow \gamma \gamma }$ in the
nonlocal case for $m_{q}=0.31-0.45$ GeV, $\Lambda =1(2)$ GeV at $M_{\sigma
}=0.44,$ $0.6$ GeV.

\vspace*{.1cm}

\begin{tabular}{|l|l|l|l|l|}
\hline
$m_{q}$ (GeV) & \ 0.31 & \ 0.35 & \ 0.40 & \ 0.45 \\ \hline
\begin{tabular}{l}
$\Gamma _{\sigma \rightarrow \gamma \gamma }$ (keV) \\ 
at $M_{\sigma }=0.44$ GeV%
\end{tabular}
& 
\begin{tabular}{l}
0.238 \\ 
(0.196)%
\end{tabular}
& 
\begin{tabular}{l}
0.192 \\ 
(0.159)%
\end{tabular}
& 
\begin{tabular}{l}
0.152 \\ 
(0.127)%
\end{tabular}
& 
\begin{tabular}{l}
0.124 \\ 
(0.105)%
\end{tabular}
\\ \hline
\begin{tabular}{l}
$\Gamma _{\sigma \rightarrow \gamma \gamma }$ (keV) \\ 
at $M_{\sigma }=0.6$ GeV%
\end{tabular}
& 
\begin{tabular}{l}
0.529 \\ 
(0.512)%
\end{tabular}
& 
\begin{tabular}{l}
0.458 \\ 
(0.415)%
\end{tabular}
& 
\begin{tabular}{l}
0.361 \\ 
(0.327)%
\end{tabular}
& 
\begin{tabular}{l}
0.294 \\ 
(0.267)%
\end{tabular}
\\ \hline
\end{tabular}
\end{center}

The decay widths decrease slowly for increasing quark mass while the
dependence on the cutoff is very weak. The numerical analysis shows that 
\begin{equation}
\Gamma _{\sigma \rightarrow \gamma \gamma }<1\text{ keV for }M_{\sigma
}<0.7-0.8\text{ GeV}\,.
\end{equation}%
Again, inclusion of the term $I_{\sigma }^{(2)}$ of Eq. (\ref{isigma2}) is
crucial to obtain these small decay widths. For instance, omission of this
term leads to the incorrect result of $\Gamma _{\sigma \rightarrow \gamma
\gamma }=1.9$ keV for values of $m_{q}=0.35$ GeV, $\Lambda =1$ GeV and $%
M_{\sigma }=0.6$ GeV, which is almost a factor $4$ larger than the correct
result of $0.458$ keV given in Table 2.

In \cite{li}, using a Coulomb-like potential, the following expression
relating the two-photon decay widths of tensor and scalar states has been
derived 
\begin{equation}
\Gamma _{\sigma \equiv \overline{n}n\rightarrow 2\gamma }(0^{++})=k\left( 
\frac{M_{N}(0^{++})}{M_{N}(2^{++})}\right) ^{m}\Gamma _{\overline{n}%
n\rightarrow 2\gamma }(2^{++})\,  \label{li}
\end{equation}%
where $m=3$. The coefficient $k$ is $15/4$ in a non-relativistic
calculation, but becomes smaller ($k\sim 2$) when considering relativistic
corrections. Choosing as input $M_{N}(2^{++})=1.275$ GeV and $\Gamma _{%
\overline{n}n\rightarrow 2\gamma }(2^{++})=2.60\pm 0.24$ keV, Eq. (\ref{li})
results with $k\sim 2$ in values of $\Gamma _{\overline{n}n\rightarrow
2\gamma }(0^{++})\sim 0.21$ and $0.54$ keV for $M_{\sigma }=0.44$ and $0.6$
GeV, respectively. These results are close to the corresponding numbers of
Table 2. As discussed in \cite{chanowitz} different values of the parameter $%
m$ are obtained for different forms of the quark-antiquark potential: for
instance, $m=-1/3$ corresponds to a linear potential. Then, in \cite%
{chanowitz} the value $m=0$ in Eq. (\ref{li}) is considered and in Ref. \cite%
{pennington} a value $m=0.3 - 1$, leading to a larger decay width, is
discussed. Here notice that our result for a light quarkonium is rather in
agreement with the choice $m=3$ and with $k\sim 2,$ see also the model in 
\cite{giacosa1} where an even smaller value of $k$ is obtained.

Notice that we have only considered sigma masses below the constituent quark
mass threshold with $M_{\sigma }<2m_{q}$ and masses $m_{q}$ in the range of
0.25 to 0.45 GeV. In order to go beyond this limit one should (i) either
increase the constituent quark mass as done in Ref.~\cite{giacosa1} where
the $\gamma \gamma $ decays of the scalars between $1$ and $1.8$ GeV have
been investigated or (ii) use more general quark propagators which include
or mimic confinement. A drawback of these extensions is that the results
reached contain a stronger model dependence, thus we do not consider these
options here.

\section{Conclusions}

In this work we use the formalism developed in Ref.~\cite{faessler} to study
the decay of a scalar quarkonium state into two photons focusing in
particular on a technical caveat of this process: a term, not present in the
usual $\pi ^{0}\rightarrow \gamma \gamma $ transition amplitude, is
responsible for a sizable suppression of the $\sigma \rightarrow \gamma
\gamma $ decay rate. We considered the process $\sigma \rightarrow \gamma
\gamma $ in the quarkonium picture both for local and nonlocal approaches.
In particular the nonlocal approach allows for a realistic treatment of the
finite size effects of the $\sigma $-meson. Similar results are obtained in
both cases for masses $M_{\sigma }$ well below the $2m_{q}$ threshold.
Closer to threshold the two-photon decay width in the local case should be
taken with great care, since the momentum dependence of the coupling
constant is not properly taken into account. Only the nonlocal result,
including finite size effects, is reliable and considerably smaller than for
the local case.

Our final result $\Gamma _{\sigma \rightarrow \gamma \gamma }<1$ keV is
smaller the results of dispersive analysis of reaction $\gamma \gamma
\rightarrow \pi ^{0}\pi ^{0}$ done in Refs.~\cite{pennington,roca}. Note,
that the framework developed in Ref.~\cite{roca} was based on approach of
Ref.~\cite{pennington}. The result of Ref.~\cite{pennington} evaluated at
the $M_{\sigma }=441$ MeV is $\Gamma _{\sigma \rightarrow \gamma \gamma
}=4.1\pm 0.3$ $\mathrm{keV}$, while the result of Ref.~\cite{roca} is around
a 40\% smaller than that in Ref.~\cite{pennington}, mainly due to a smaller $%
\sigma \pi \pi $ coupling.

When discussing our results it is important to stress that two aspects have
not been considered. The first one is the possible role of pion loops. Note,
that we consider a scenario where the $\sigma $ meson is a pure $\bar{q}q$
Fock state and, therefore, the $\sigma $ couples directly to its
constituents -- quarks. The coupling with other mesons (e.g. pions) goes via
quark loops (a direct coupling of the $\sigma $ to pions is not present).
Inclusion in a such picture of pion loops generating $\sigma \rightarrow
\gamma \gamma $ transition can occur as in Fig. 5: the corresponding
amplitude is suppressed of a factor $1/N_{c}$. Our framework is restricted
to the one-loop approximation and to the dominant term(s) in the $1/N_{c}$
expansion. However, being in Nature $N_{c}=3$ an explicit calculation of the
next-to-leading order would surely be helpful to quantify its contribution
but goes beyond the scope of present paper and is left as outlook. Notice
that, if we propose that the $\sigma $ meson is not pure $\bar{q}q$ state
and there is also two-pion component contribution to the $\sigma $ meson
Fock state, then we should include both possible intermediate states $\bar{q}%
q$ and $2\pi $ contributing to the two-photon transition of the $\sigma $.
We plan to study the second scenario - $\sigma $ being mixture of $\bar{q}q$
and $2\pi $ - in future.

\begin{figure}[btbp]
\centering{\ }
\par
\vspace*{-.5cm} \epsfig{figure=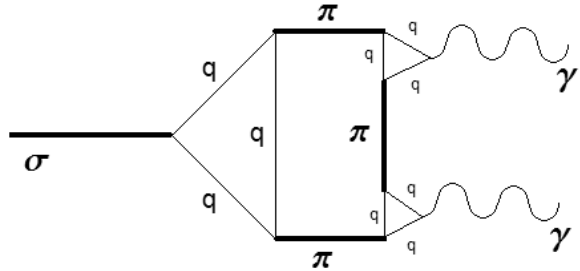,scale=.5} \vspace*{-.35cm}
\caption{Pion-loop contribution to the two-photon decay within our model(s).
Thik lines refer to mesons, thin lines to quarks.}
\end{figure}

The second aspect is related to the inclusion of finite-width effects in the
evaluation of the full $\gamma \gamma $-width of the sigma resonance. A
careful description of this point would require the precise knowledge of the
propagator of the sigma meson dressed by pion clouds: in such a way a
definition of the spectral function allowing to integrate over the whole
mass range up to $1.27$ GeV is possible \cite{pagliara}. Thus, also this
aspect is related to pion loops and is not performed here. However, using
trial distributions such as Breit-Wigner one and the generalized form of
Ref. \cite{pagliara} and varying the mass and the width an increase of few
percent is observed. For instance, be the decay rate $\Gamma _{\sigma
\rightarrow \gamma \gamma }=0.458$ $\mathrm{keV}$ at $M_{\sigma }=0.6$ GeV
as in the second column of Table 2: considering a 500 MeV wide Breit-Wigner
distribution the integrated width $\Gamma _{\sigma \rightarrow \gamma \gamma
}$ reads $0.66$ $\mathrm{keV}$. While such effects are surely important in a
precision study of the two-photon decay width they does not change the
qualitative outcome of the present paper. The corresponding integrated
signals decay width(s) reported by \cite{pdg} are $\Gamma _{\sigma
\rightarrow \gamma \gamma }=3.8\pm 1.5$ $\mathrm{keV}$ and $5.4\pm 2.3$ keV,
values which are not accepted as average or fit. Note that a confirmation of
a large experimental value, contrary to usual belief, does not favor a
quarkonium interpretation of the sigma meson. As noted in the PDG2000 \cite%
{pdg2000}, the large value for $\Gamma _{\sigma \rightarrow \gamma \gamma }$
could arise from an additional contribution of the broad $f_{0}(1370)$. A
clear experimental determination of the two-photon decay width would
certainly help in clarifying the discussion related to the nature of the $%
\sigma $-meson.

\begin{acknowledgments}
This work was supported by the DFG under contracts FA67/31-1 and GRK683.
This research is also part of the EU Integrated Infrastructure Initiative
Hadronphysics project under contract number RII3-CT-2004-506078 and
President grant of Russia "Scientific Schools" No. 5103.2006.2.
\end{acknowledgments}

\end{document}